# Dynamics and rheology of 2D colloidal crystals with active anisotropic impurities

Jacob John[1] and Giovanniantonio Natale[1]

[1]Department of Chemical and Petroleum Engineering, University of Calgary, Alberta, Canada, T2N1N4.

Email: gnatale@ucalgary.ca

**Abstract**

Active motion at complex fluid-fluid interfaces is a ubiquitous phenomenon in nature. However, an intriguing question that is not fully addressed is how active motion affects and gets influenced by its complex environment. Here, we design a 2D colloidal crystal containing active particles as a model system to understand the mechanics of complex interfaces in the presence of activity. We characterize the dynamics, rheology and phase behavior of 2D colloidal crystals formed using spherical polystyrene (PS) particles in the presence of active PS-Platinum Janus particles. In the presence of activity, the overall crystal becomes dynamic with a heterogeneous spatial distribution of disorder. Through particle-tracking microrheology and interfacial shear rheology, we report the discovery of a phase transition from solid-like to liquid-like interface driven by the active motion of a small number of active impurities in the crystal. The local perturbations induced by the active impurities have long-range effects on the dynamics of particles in 2D colloidal crystals modifying their overall viscoelasticity. The correlations between microstructure and dynamics from our experiments can provide insights into the behavior of a broad range of complex systems such as, motion of particles at oil-water interfaces in Pickering emulsion microreactors and the motion of molecular motors on cell membranes.

Materials with disorder in their microstructure are ubiquitous and they are present in foods [1], rocks [2], concrete and several biological tissues [3]. Unlike crystalline solids, the yielding behavior of disordered materials is catastrophic and is often difficult to predict. While the occurrence of ductile flow is prevalent in crystalline materials at yield point, defined as the value of stress beyond which a material undergoes a transition from elastic deformation to viscous flow [4], the propagation of localized defects is rapid during yielding in the case of disordered materials [5]. One of the reasons for unpredictability of the mechanics of disordered solids is the lack of microstructural understanding about their mechanics at different length scales. Therefore, it is important to understand the role of localized perturbations and defects which could act as precursors to failure in disordered materials.

2D colloidal crystals which are formed due to self-assembly of charged particles at fluid-fluid interfaces [6], is an excellent model system for studying the effect of disorder and local perturbations in materials. The interparticle repulsion due to charges present on the surface of the particles force the spontaneous organization of the particles into a well-structured crystal-like arrangement. By introducing active anisotropic colloids as impurities in 2D colloidal crystals, the strength of the perturbations and disorder can be controlled. Moreover, the microstructural changes in the crystal induced due to activity can be easily visualized using a microscope. Active colloids, both living and synthetic, use energy from their environment to undergo selfpropulsion[7], exhibiting interesting dynamics different from that of Brownian motion [8, 9]. Active motion is also a ubiquitous phenomenon at fluid-fluid interfaces in nature [10, 11]. Bacterial cells in biofilms [12], particles at the interface of Pickering emulsion microreactors [13], protein motor molecules at cell membranes [14] and human crowds [15] are all examples of active motion at complex interfaces. As an experimental system, 2D colloidal crystals with active colloidal particles contain the essential physics to unravel how the self-propulsion of active particles affects and is affected by its complex environment.

The presence of anisotropic impurities in 2D colloidal crystals is known to affect the crystal microstructure. The addition of ellipsoidal polystyrene (PS) particles as impurities in 2D colloidal crystals made of charged spherical PS particles can reduce its crystallinity also separate the crystal into smaller domains [16]. The presence of magnetic Janus particles in 2D colloidal crystals has shown the ability to get transported within the colloidal crystal [17]. The active particles either behave as interstitials, i.e., moving around particles within the crystal, or can act as active atoms by becoming part of the colloidal crystal depending on the orientation of the Janus particles at the fluid-fluid interface [18]. These findings point to the importance of the dynamics of individual active particles on the origin of different phases in active-passive particle mixtures. Different modes of phase transitions have been observed in earlier mixtures of active and passive particles specifically in bulk mixtures. Some of these include crystallization from a gas-like state [19], recrystallization by removal of grain boundaries and melting [20]. At high ratios of active particles, the existence of liquid-crystal, phase-separated crystals and gas-like phases have been also observed in mixtures of active and passive particles [21, 22, 23]. In such mixtures, the phase transition from a homogeneous to a phase-separated state can occur either due to a phenomenon called motility-induced phase separation (MIPS), where passive particles tend to aggregate at locations where their movement is slow [24] or due to depletion interaction between the passive particles. While there are theories namely the Kosterlitz-Thouless-Halperin-Nelson-Young (KTHNY) theory predicting a two-step melting process and the grain boundary mediated one-step melting induced due to temperature in defect-free 2D colloidal crystals, there is a lack of understanding about phase transitions induced by active colloids and in the presence of defects.[25]

Here we explore the behaviour of 2D colloidal crystal in the presence of active impurities. We investigate how local perturbations induced due to the active motion of Janus particles can affect the large-scale dynamics of 2D colloidal crystals. We show how the presence of activity at low active-to-passive particle ratios can give rise to a phase transition from viscoelastic solid to liquid-like interface. We elucidate, by means of interfacial micro and shear rheology, how the competition between repulsive interactions between particles and the hydrodynamic propagation of disorder gives rise to yielding behavior in colloidal crystals with active impurities.

## Colloidal crystals with Janus impurities in the absence of activity

The presence of Janus particles as impurities in the 2D colloidal crystal affects its structure and rheology even in the absence of activity (0% $H_2O_2$). The micrographs in Figure 1 shows particle organization at the interface in the presence of active particles with ρ = 1:320 and different surface densities (φ). The presence of Janus impurities introduces imperfections in the colloidal crystals in the form of point defects, dislocations, and grain boundaries. This is evident in the corresponding Fast Fourier Transforms (FFT) of the micrographs shown in Figure 1. FFT images of the micrographs represent the spatial distribution of crystal periodicity, shown using bright spots and patterns.

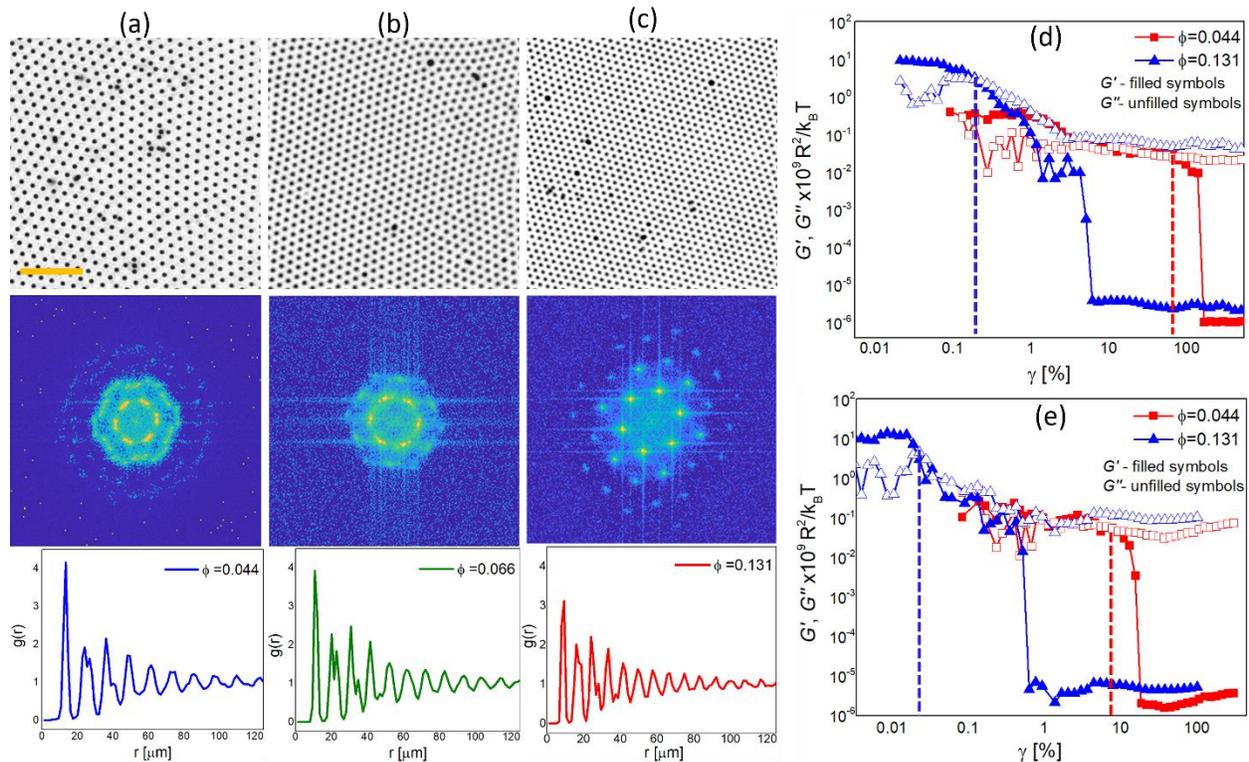

**Figure 1: Microstructure and rheology in the absence of activity**. Micrographs showing the organization of PS particles at the oil-water interface in the presence of active particles (ρ = 1:320), their corresponding FFT images and radial distribution function with (a) φ=0.044 (b) φ=0.066 and (c) φ=0.131. The scale bar shown is 25µm. Variation of interfacial storage modulus (G') and Loss modulus (G'') as a function of shear strain for colloidal crystals (e) without active particles and (e) with (ρ = 1:320), at 0% $H_2O_2$.

The FFT of the micrographs shows variation in patterns at different surface densities. The diffuse halo in the case of φ=0.044 changes to sharp spots at φ=0.131. This indicates that at φ=0.044, the presence of Janus particle leads to the formation of a disordered crystal while at φ=0.131, a well-ordered crystal with hexagonal symmetry is formed. The formation of defects due to Janus impurities is more pronounced in the case of the colloidal crystal with a lower surface density. To quantitatively analyze the ordering of the particles shown in the micrographs in Figure 1, we calculated the radial distribution function g(r) within the area of the image. For a well-ordered crystal, g(r) will show peaks at distances that at multiples of lattice constants for a broad range of r values, indicating a long-range order [26]. As shown in Figure 1, the number of peaks is higher in the case of the crystal with higher surface density. Therefore, at the same ratio of Janus particles (ρ), a lesser number of defects are created in the case of the crystal with higher surface density.

Figures 1 (d) and (e) respectively show the amplitude sweep response of the interface in the absence and presence of Janus particle impurities in the colloidal crystals, for two different surface densities φ=0.044 and φ=0.131. As shown in Figure 1 (d), at small strains, both φ=0.044 and φ=0.131 have a plateau value of G' and the G' is greater than G''. This indicates that the 2D colloidal crystal interface exhibits a viscoelastic solid-like behavior. As expected, the interface with higher surface density (φ=0.131) has a higher value of plateau G' compared to that of φ=0.044, suggesting the higher elasticity of the interface with φ=0.131. The elastic response of the 2D colloidal crystal interface could be arising from the solid-like crystalline arrangement of the particles at the interface. Fluid-fluid interfaces containing, silver nanoparticles [27], monodisperse rod-like particles [28] and spherical polystyrene particles are known to exhibit a gel-like behavior due to interfacial jamming of particles and network formation. At higher strain amplitudes, the G' shows a decrease, indicating a shear-thinning regime and a rupture of the crystalline microstructure. The dashed lines indicate the yield strain at which G' becomes lower than G''. As seen in Figure 1 (d), the yield strain is lower in the case of φ=0.131, indicating that the colloidal crystal is more brittle at higher surface density.

In the presence of Janus particles, the yield strain decreases by an order of magnitude, in both the cases, φ=0.044 and φ=0.131, as shown in the Figure 1 (e). The defects in the crystal introduced due to the presence of Janus particle impurities (Figure S3), causes the solid to liquid transition under shear to occur at a lower strain amplitude. 3D colloidal crystals are known to undergo yielding under shear deformation by breaking apart into smaller regions of crystals [29]. Observations of series of bond breaking events have also been made in the case of 2D colloidal gels under shear [30].

## Propagation of defects in the presence of activity.

Figure 2 shows snapshots of the microstructure and the corresponding Voronoi diagrams of the 2D colloidal crystals with Janus particles in presence of activity. The changes in the Voronoi diagram at different instances in time shows how the defects propagate in the crystal in the presence of activity. The three panels compare the effect of surface density at the same active particle ratio and $H_2O_2$ concentration. The presence of activity causes the active particles to move within a hexagonal lattice of the colloidal crystal. The number of defects are highest in the case of φ=0.044. As time progresses, the location of the defects moves within the crystal. This can be observed as changes in the pattern of the green and blue cells in the Voronoi diagrams in Figure 2 (a). With increase in elasticity of the colloidal crystal at higher surface density, mobility of the active particles in the crystal reduces. The number of defects also reduces with increase in surface density, as can be seen in the Voronoi diagrams in Figures 2(b) and (c). The time-lapse Voronoi images show that in the case of φ=0.044, the motility of the active particles cause the defects to form a string-like pattern. Such a pattern is formed during the diffusion of defects in the crystals [31]. In 3D colloidal crystals, the motion of defects is a mode of dissipation of stresses within the crystal[32]. The microstructure of the 2D colloidal crystals at φ=0.044 and ρ=1:320 at different peroxide concentrations (Figure S4) indicate the differences in motility of the active particles. The number of defects and their motion also increases. In the case of 2D colloidal crystals formed using polystyrene particles, it has been shown that the motion of defects begins when point defects dissociate into dislocation pairs [33].

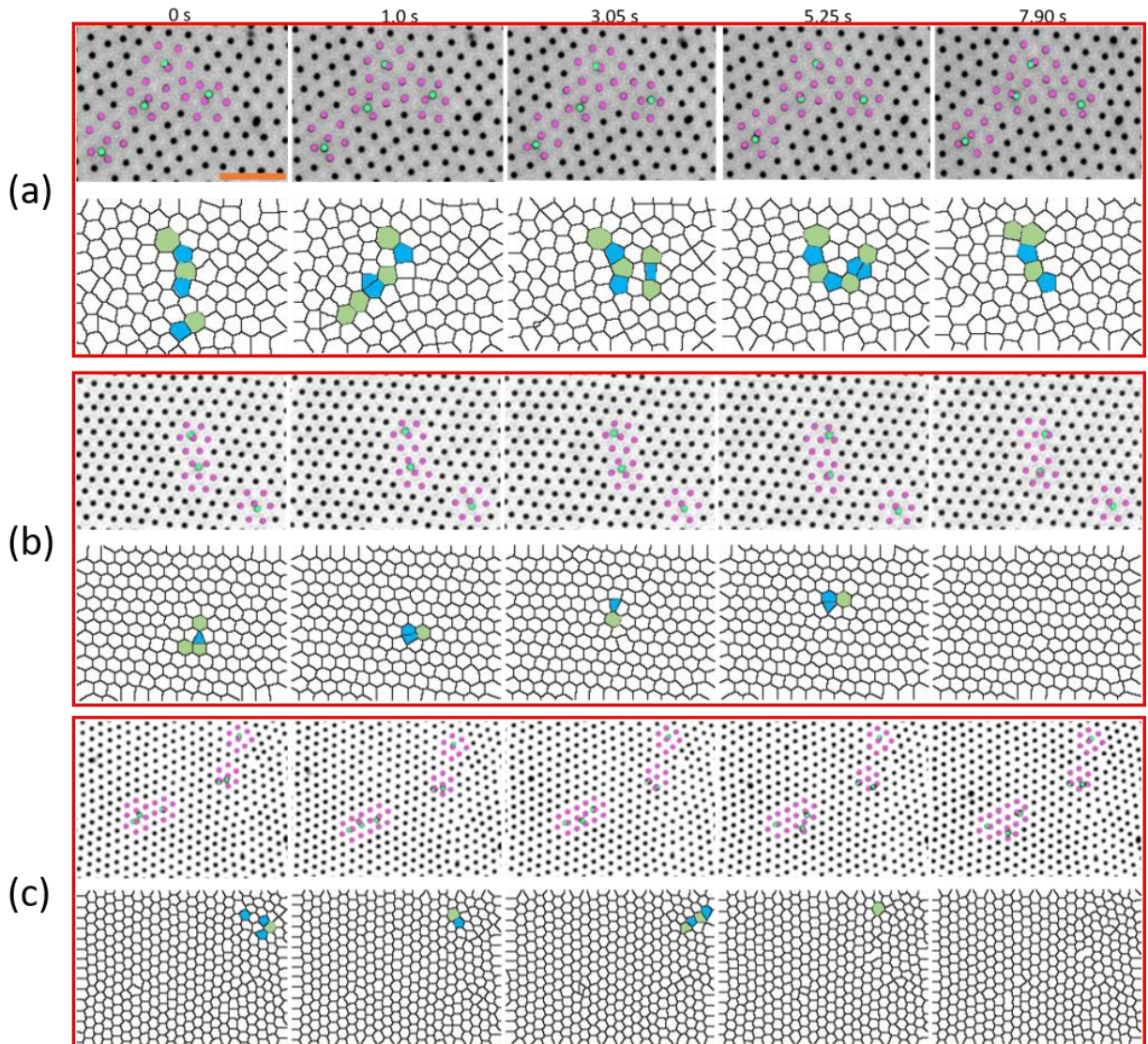

**Figure 2: Microstructure in the presence of activity.** Snapshots of particle microstructure and their corresponding Voronoi plots at different time intervals for (a) $\varphi=0.044$ (b) $\varphi=0.066$ and (c) $\varphi=0.131$ at $\rho=1:320$ and $H_2O_2$ concentration of 3%. The scale bar shown is 25 µm. The active particles are indicated using green color and their neighboring passive particles using pink color to locate the changes in the position of the particles. The corresponding Voronoi plots indicate the location of defects in the colloidal crystal. Defects are identified as the regions where the number of nearest neighbors around particles are not equal to six. Cells with seven edges are filled with green colors and the cells with five edges are filled with blue color.

It is observed that with increasing surface density ($\varphi$), the slope of the MSDs, $\alpha$ reduces towards zero, indicating that the motion of the particles are restricted (Figure S5). In the presence of activity, the particles exhibit greater dynamics with slopes close to 1, while the slope remains close to 0 in the absence of activity (0% $H_2O_2$). The MSDs shown in Figure S5 are only an average representation of the dynamics of the interface. However, the Voronoi diagrams shown in Figure 2 indicate that the defects are heterogeneously distributed in the colloidal crystal.

**Spatial variation of dynamics: Heterogeneity.**
In order to quantify the heterogeneity in dynamics, the slopes, α, from individual MSDs of all the particles in the colloidal crystal are represented on 2D maps. On the maps, we report the α values of the MSDs of individual particles using Voronoi cells at the corresponding positions of the particles in the colloidal crystal. The individual cells in the Voronoi diagram were color coded based on the α value of the particle at the position. The heterogeneity map provides a visual description of the spatial variation of mechanical properties of the 2D colloidal crystals in the presence of activity. Figures 3 (a), (b) and (c) show the heterogeneity map for three different surface densities (φ) at the highest active particle ratio (ρ=1:320) and $H_2O_2$ concentration (15%). As shown in Figure 3 (a), at φ=0.044, majority of the cells have a color representing α>1 indicating the larger displacement of the particles in the colloidal crystal. The long-range effects result from the ability of the defects to propagate larger lengths in the colloidal crystal. With increase in φ, the color of the Voronoi cells in the map changes to represent lower values of α and regions with α>1 appear constraint within the crystal in the case of φ=0.131.

In Figures 3 (d), (e) and (f), the heterogeneity maps show the effect of activity ($H_2O_2$ concentration) on the dynamics of the particles in the colloidal crystals at φ=0.066. The α values on the map indicates that the interface has a viscoelastic liquid-like behavior at 3% and 15% $H_2O_2$ concentrations. The maps also indicate that heterogeneous regions with higher α values, of the order of tens of micrometers exist in the interface due to activity of the Janus particles. These mechanically weak regions could reduce the elasticity of the 2D colloidal crystal. The average values of α for all the particles in the 2D colloidal crystal is represented in the 3D plot shown in Figure 3 (g) for all peroxide concentrations, surface densities and active particle ratios used in this work. The phase diagram clearly indicates a phase transition where the dynamics of the colloidal crystal changes from sub-diffusive (α<1) to super-diffusive (α<1) with variation in composition and activity in the 2D colloidal crystal. Significant changes in characteristics can be observed in the case of lower surface densities, such as φ=0.044 and φ=0.066. With increase in activity ($H_2O_2$ concentration), the α increases to values greater than 1. The contour plot shown in Figure 3 (h) for ρ=1:320 clearly shows the phase boundary when activity and surface density are varied. The super-diffusive nature occurs when the interface acquires a liquid-like nature due to the loss of elasticity in presence of active motion of the Janus particles. The changes in viscoelastic properties of the 2D colloidal crystal due to active motion of the Janus particles can be understood using interfacial shear rheology.

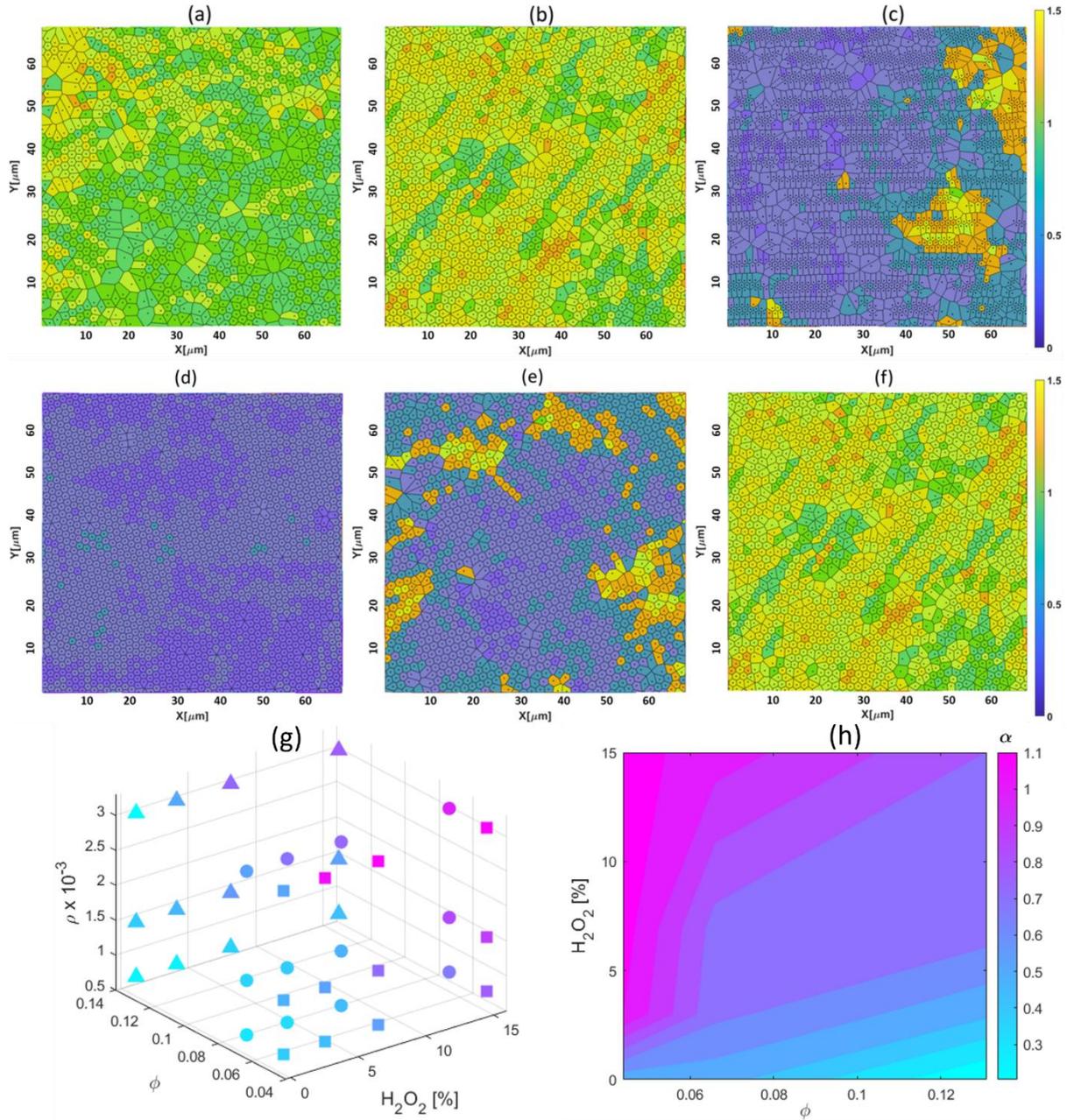

**Figure 3: Heterogeneous dynamics and phase diagram** Top panel: Heterogeneity maps showing regions with different α values at 15% $H_2O_2$ and ρ=1:320 for (a) φ=0.044 (b) φ=0.066 and (c) φ=0.131. Second panel: Heterogeneity maps for φ=0.066, ρ=1:320 and different $H_2O_2$ concentrations, (d) 0% (e) 3% and (f) 15%. Phase diagram: (g) The average values of α for MSDs of all particles in the colloidal crystal for different compositions of the interface studied in this work, represented using a 3D phase diagram. (h) 2D contour plots showing the effect of $H_2O_2$ concentration and surface density (φ), for ρ=1:320.

## Activity controls melting of the colloidal crystal

Figure 4 (a) shows the variation of G' and G'' as a function of strain amplitude for 2D colloidal crystals with φ=0.131 and different $H_2O_2$ concentrations. In the absence of activity, the 2D colloidal crystal exhibits a viscoelastic solid-like interface, with G' exhibiting a plateau and G' > G'' at small strain amplitudes. The viscoelastic solid-like nature of the interface is also evident in the frequency sweep response showing a frequency independent nature and G'>G'' for three decades of ω. However, with increase in $H_2O_2$ concentration, the plateau value of G' in the linear regime reduces, accompanied by a decrease in the yield strain. This indicates a loss of elasticity of the 2D colloidal crystal. In the case of 15% $H_2O_2$ concentration, the interface with 2D colloidal crystal becomes so weak that the plateau region of G' and G'' is undetectable at small strain amplitudes. This happens due to low torque values measured due to the loss of elasticity of the interface due to activity. Similar response can be observed also in the frequency sweep behavior of the interface, shown in Figure 4 (b). The interfacial shear rheology experiments complement the results obtained from the microrheology experiments, which indicate a solid to liquid melting transition induced due to activity of a small number of active particles.

Although melting phase transition of 2D colloidal crystals has been shown to occur for a variety of external stimuli such as temperature[34], magnetic field[35] and light[36], this is, to our knowledge, the first ever report of melting induced due to active colloids. In all the later cases, the solid-liquid phase transition was observed to align with the predictions of the KTHNY theory. To obtain insights into the mechanism of the phase transition induced by the active particles, we probe the dynamics of the particles in the crystal using two-point (2P) microrheology. We measure two types of spatially correlated displacements ($D_{rr}$ and $D_{\theta\theta}$) of pairs of particles in the crystal. $D_{rr}$ and $D_{\theta\theta}$ are both functions of R and τ and represents the correlated displacement along and perpendicular to the line joining two particles, respectively.

Figures 4 (c) and (d) respectively show the variation of $D_{rr}$ and $D_{\theta\theta}$ as a function of distance R for the composition which shows the highest effect of activity on the crystal. At all the separation distances, both the correlated displacements exhibit a behavior independent of R. Thus, the correlated displacements between two particles which are near and far are equal. This behavior of non-decaying correlated motion indicates that the deformation field is persistent over long range and arises from the viscoelastic nature of the interface induced due to the dipolar interactions of the particles in the crystal [37]. The close proximity of the particles additionally contributes to interactions between the deformation fields. This behavior is very different from the case of homogeneous three-dimensional systems where the correlations $D_{rr}$ and $D_{\theta\theta}$ are known to decay as 1/R and $1/R^2$ respectively [38]. The variation of $D_{rr}$ and $D_{\theta\theta}$ as a function of time exhibits even more interesting behavior. Even at a distance of R=100 μm, $D_{rr}$ and $D_{\theta\theta}$ exhibits a time-independent behavior (Figure S7, (a) and (b)), indicating that the momentum diffusion at the interface occurs faster than 0.05 s (inverse of the frame-rate used).

Additionally, both the spatial correlation functions $D_{rr}$ and $D_{\theta\theta}$ exhibit fluctuations, which arises from the heterogeneity in dynamics of the particles in the crystal, also shown using Voronoi diagrams in Figure 3. In absence of activity, the fluctuations are low and the $D_{rr}$ and $D_{\theta\theta}$ values are one order lower in magnitude (Figure S8 (c) and (d)). In presence of activity, both $D_{rr}$ and $D_{\theta\theta}$ are dependent on the lag time (Figure S8 (a) and (b)). The correlated displacements increase with increase in the lag time used for calculating the $D_{rr}$ and $D_{\theta\theta}$. This behavior arises form the viscoelastic liquid-like behavior of the interface because of melting of the crystal, induced due to activity [37]. However, in the case of $\varphi=0.131$ and 0% $H_2O_2$, $D_{rr}$ and $D_{\theta\theta}$ are independent of the lag time, (Figure S8 (c) and (d)), indicating that the elastic behavior of the interface is retained.

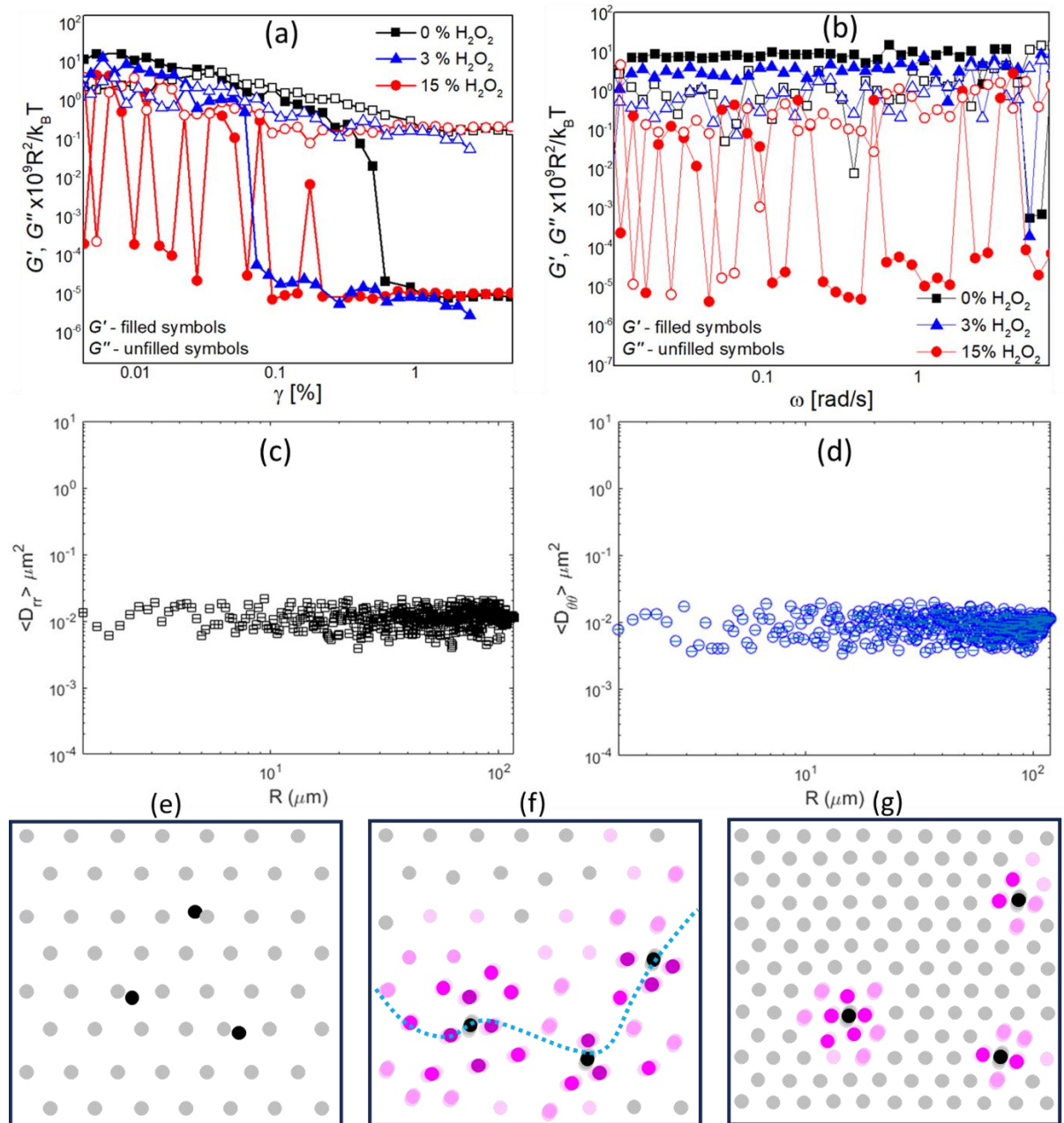

**Figure 4: Activity controls phase transition and mechanics:** Variation of (G') and (G'') as a function of (a) shear strain and (b) frequency for $\rho=1:320$ and $\varphi=0.131$, at different $H_2O_2$ concentrations. Variation of (c) $D_{rr}$ and (d) $D_{\theta\theta}$ as a function of R for a lag time ($\tau$) of 0.1 s for $\varphi=0.044$, $\rho=1:320$ and 15% $H_2O_2$. Schematic representation of the general mechanism under different situations such as, (e) Low $\varphi$ and no activity, where the crystal structure is unaffected. (f) Low $\varphi$ and high activity, where the motion of the active particles perturbs the neighboring passive particles resulting in the creation of a string of defects (dashed line). (g) High $\varphi$ and high activity, where the perturbations are contained locally. The active particles are represented using black symbols. Higher intensity of the pink color indicates higher motility of the passive particles.

As shown in the time lapse of the microstructure in Figure 2 and Figure S3, the active particles does not navigate through the entire colloidal crystal. This indicates that the electrostatic coupling between the active and passive particles are dominant [18]. Active particles can get trapped close to the passive particle, occasionally orbiting around a passive particle and also moving towards a neighbouring passive particle. Undulations in contact line created by the particles promote capillary attraction between the active and passive particles (with a potential of magnitude ~50 kBT). These undulations also cause changes in area of the Pt hemisphere exposed to $H_2O_2$ in the subphase [39]. This results in either the particle escaping the trap and moving to a neighbouring particle or orbiting around the same passive particle. It has been reported earlier that gradients in fuel and hydrodynamic coupling contributes to the orbiting of active particles around obstacles [40, 41]. Both these events give rise to fluctuations of passive particles from their equilibrium positions in the colloidal crystal. This leads to an overall increase in motility of the neighbouring passive particles in proximity of the active particle in the case of low particle densities.

The motion of Janus particles results in the creation of string-like defects shown schematically in Figure 4 (f). This can also be observed in the microstructure and Voronoi diagrams shown in Figure 2 and Figure S3. The propagation of these chain-like defects causes the dissociation of the colloidal crystal into polycrystalline domains with short-ranged order, which results in decrease in elasticity of the crystal. However, in the case of densely packed crystal, the dipolar repulsive interactions between the passive particles dominate and keep the perturbations caused by active particles contained locally, as shown in the phase diagram in Figure 3 (h) for the case of $\varphi=0.131$. These observations suggest the first-order grain boundary mediated melting as the phase transition mechanism for activity-induced meting of 2D colloidal crystals.

In summary, the role of active anisotropic particle as impurities in 2D colloidal crystals is explored using microrheology and interfacial shear rheology. The presence of Janus particles as impurities can introduces defects in the 2D colloidal crystals even in the absence of activity. Further, the defects also dampen the elasticity of the colloidal crystal. The motion of active particles propagate the defects along the 2D colloidal crystal. This creates heterogeneity in dynamics of the particles in the 2D colloidal crystal and results in melting phase transition in the case of low surface densities and high activity. The effect of surface density, active-passive particle ratio and activity on the phase behavior of the 2D colloidal crystal is explored and used to build a 3D phase diagram for these systems. Two-point microrheology shows the persistence of long-range correlated displacements, independent of time and and distance in the presence of active motion pointing to the dominant effect of electrostatic interactions. Our results show that active colloids can be used to realize new interfacial systems whose mechanical properties can be modified on demand via the discovered active melting mechanism.

**Methods**

Materials Sulphate functionalized, negatively charged, spherical Polystyrene (PS) particles (1 μm dia), used in this work was purchased from ThermoFisher Scientific. It was supplied as an aqueous suspension of 8% w/v. Janus particles used in this work was synthesized using the same 1 μm PS particles by sputter coating one hemisphere of the particles with 20 nm of platinum. The details of the method followed for the synthesis of Janus particles used in this work are given in the supporting information. In order to create oil-water interfaces, 30% glycerol in deionized Milli-Q water was used as the subphase and decane was used the top phase. Decane (>99%, AR grade) was supplied by ThermoFisher Scientific. Hydrogen peroxide (30%) was purchased from ThermoFisher Scientific. Experiments 2D colloidal crystals were created by dispersing spherical Polystyrene (PS) particles (1 μm dia) on to a clean oil-water interface. PS-Platinum Janus particles were used as active impurities in the colloidal crystal. The dynamics of the particles in the 2D colloidal crystal was studied using optical microscopy. For the visualization, a custom built set up consisting of a 8 mm diameter cylinder was used (described in Figure S1). The dispersion containing either PS particles or mixtures of PS and Janus particles was dispensed carefully to the interface of the two liquids using a micro pipette. The volume of the dispersion is varied based on the desired surface density ($\varphi$) required. $\varphi$ is defined as the number of particles per $\mu m^2$ area. The motion of the particles in the 2D colloidal crystals were tracked by recording movies of the interface at 20 frames/s using a 20x objective of Leica (DMi8) microscope. The trajectories of individual particles are determined using the image processing software Image J. The characterization of the images are performed by determining its fast fourier transform, voronoi plots and radial distribution function of the images are also determined using the built-in macros in Image J. The trajectories of the particles were analyzed using an open access MATLAB code developed by Croker et al [42]. to determine the mean square displacement of the particles.

Interfacial shear rheology experiments on the 2D colloidal crystals containing active colloidal particles was performed on an Anton Paar (MCR 702) rheometer using a du Nouy ring geometry. Oil-water interface is first created before dispersing the particles at the interface. Next, the du nouy ring geometry is placed at the interface. Oscillatory shear rheology of the 2D colloidal crystals with active impurities were carried out by applying a sinusoidal deformation varying either the strain amplitude or the angular frequency and maintaining the other constant. Strain sweep experiments were preformed by varying shear strain from 0.001 to 100 %. Frequency sweep tests were performed by varying angular frequency ($\omega$) from 0.01 to 10 rad/s. The in-phase component of the output stress and the out-of-phase component of the output stress can be used to determine material properties such as the interfacial storage modulus (G') and the interfacial loss modulus (G''), respectively. G' and G'' represent the solid-like and liquid-like characteristics of the viscoelastic interface containing the 2D colloidal crystal. The amplitude sweep response shows the variation of G' and G'' as a function of strain amplitude and therefore indicates the variation of viscoelastic properties of the interface with increasing length scales of deformation. Similarly frequency sweep measurements show the variation of G' and G'' as a function of angular frequency and indicates the variation of viscoelastic properties at different time scales. It should also be noted that colloidal crystals with $\varphi$ <0.131 in the presence of activity had low torque values and therefore the poor signal-noise ratio made viscoelasticity measurements using interfacial shear rheology difficult.

**Acknowledgments**
The authors acknowledge the NSERC Discovery Grant (RGPIN-2017- 03783) and the Canada First Research Excellence Fund (CFREF) for the financial support.We also acknowledge the support of John R. Evans Leaders Fund (JELF), Canada Foundation of Innovation (Fund Number 37984).

# Supporting Information

Details of the experimental setup

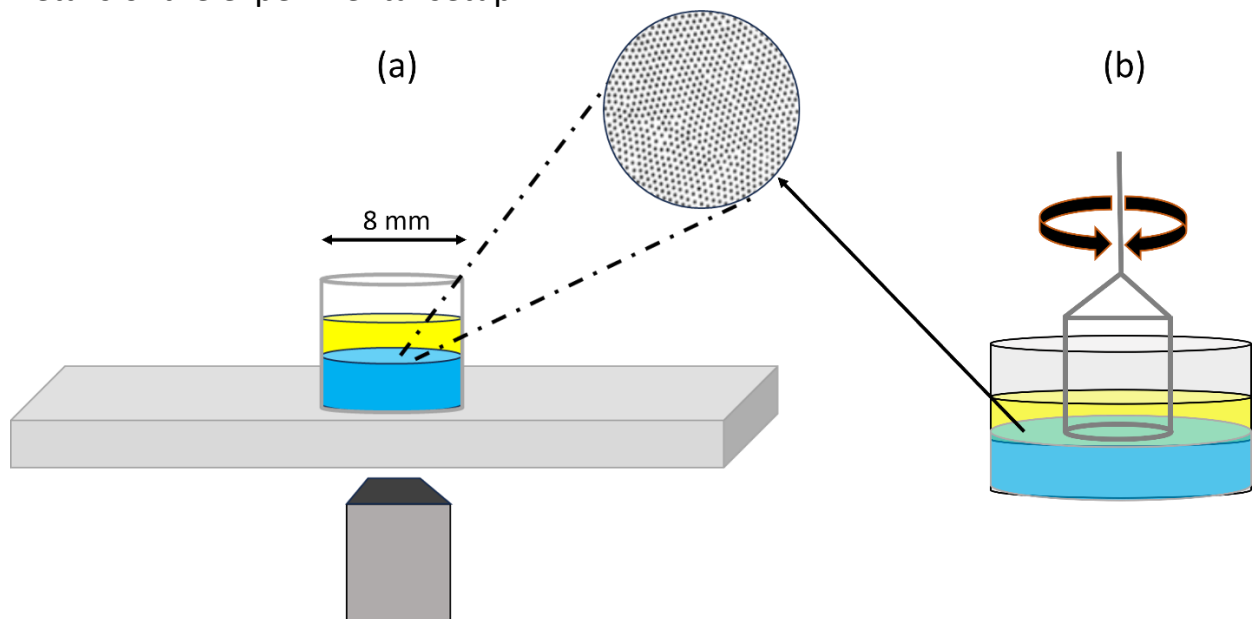

Figure S1: Schematic of the experimental set up used for (a) microscopy observation and (b) interfacial shear rheology.

Interfacial microscopy experiments were carried out using a simple homemade setup schematically shown in Figure S2 (a). The set up consists of a cylindrical raschig ring (8 mm dia) glued to a glass slide. The cylindrical well is used to create the oil-water interface where the active and passive particles are dispersed. The sub-phase used is 30% w/w glycerol in water and the super-phase is decane. The interface is first created and then the dispersion containing mixtures of passive (PS) and Janus (PS-Pt) particles are spread at the interface. The dispersion also contains ethanol in 1:2 ratio. Ethanol helps in spreading of the particles at the interface. It was observed that the particles instantaneously arrange into a 2D colloidal crystal structure. The motion of the particles in the 2D colloidal crystals were tracked by recording movies of the interface at 20 frames/s using a Leica (DMi8) microscope. A drift correction of the trajectories was carried out according to the protocol of Crocker et. al.,[1] before further analysis of the trajectories.

## Janus particles prepared by sputter coating

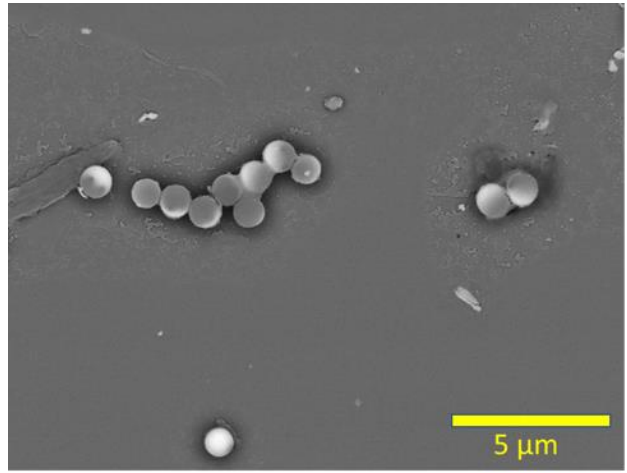

Figure S2: SEM image showing the Polystyrene-platinum Janus particles prepared by sputter coating for using as active particles in this work.

Janus particles used in this work were prepared by sputter coating method. A dispersion (0.05 wt%) containing spherical Polystyrene particles were spin coated on a glass substrate. One hemisphere of the particles were sputter coated with an initial layer (~5 nm) of titanium (Ti) for proper adherence of platinum. This is followed by the deposition of a layer (~30 nm) of platinum (Pt) using Kurt J. Lesker CMS-18 mult-source sputtering system. To obtain a dispersion of the Janus particles in water, the glass slides were sonicated in water for around 5 minutes to detach the particles from the slides.

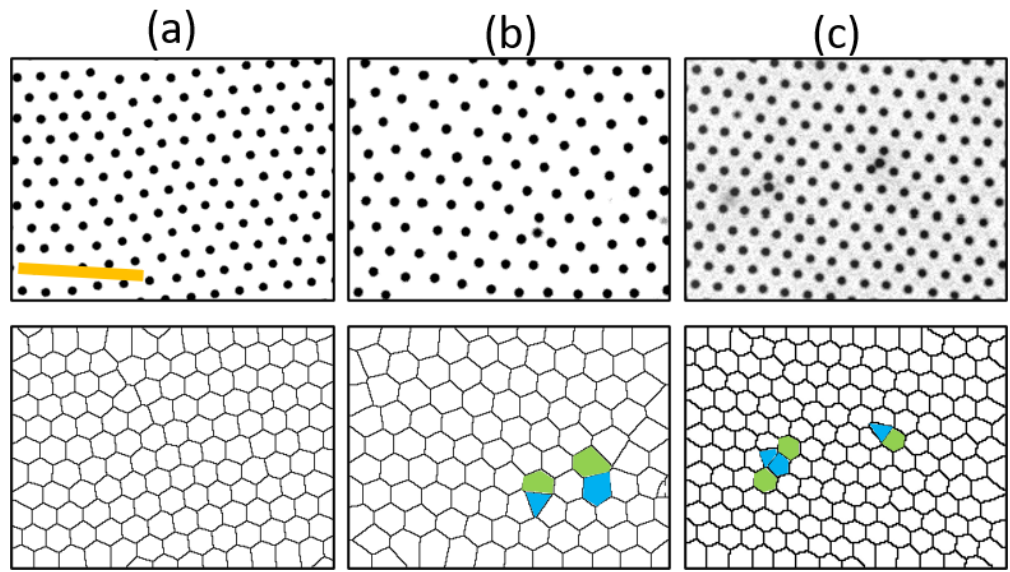

Figure S3: Micrographs showing the microstructure, and theirs corresponding Voronoi plots for colloidal crystals with φ=0.044, in the presence of active particles at different ratios, (a) ρ = 1:1280, (a) ρ = 1:640, (a) ρ = 1:320, at 0% $H_2O_2$. Scale bar shown is 25 µm.

The location of the defects in the crystals can be identified using Voronoi-tessellations of the micrographs. Voronoi diagrams represent the interface containing the particles by dividing it into regions/cells containing individual particles. The boundaries of the Voronoi cells are drawn in such a way that all points within a cell are closer to the particle in the cell than any neighboring particle. Figure S3 shows the Voronoi diagrams for the colloidal crystals with different ratios of Janus particles (ρ), and lowest surface density (φ=0.044) in the absence of activity (0% $H_2O_2$). The cells in the Voronoi plots of the micrographs have a hexagonal shape indicating that each particles have six neighbours. However, With the increase in the ratio of Janus particles, cells with more or less than six neighbours can be found. These are point defects within the colloidal crystal. Cells with seven edges are filled with green colors and the cells with five edges are filled with blue colors in figure S3. Although the Voronoi cells remain hexagonal in shape, the number of defects increases with an increase in the ratio of Janus particles. The defects are located at the position of the Janus particles in the colloidal crystal.

## Microstructure-dynamics of defects

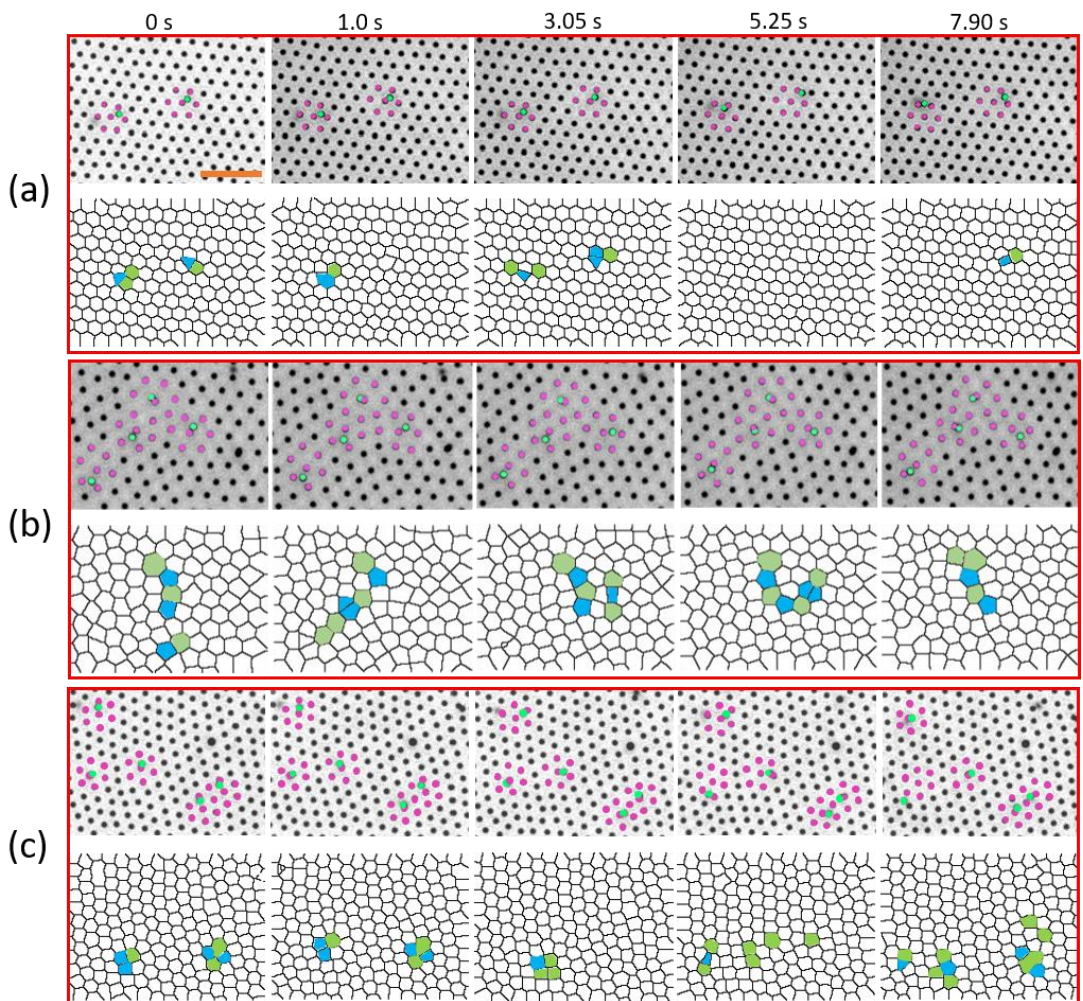

Figure S4: Snapshots of particle microstructure and their corresponding Voronoi plots at different time intervals for φ=0.044 and ρ=1:320 at (a) 0% (b) 3% and (C) 15% $H_2O_2$ concentrations. The scale bar shown is 25 µm.

The time-lapse of the micrographs at different peroxide concentrations indicate the differences in motility of the active particles in the 2D colloidal crystals. As expected, with increase in the peroxide concentration, the mobility of the interstitial active particles increases. Simultaneously, the number of defects and their motion also increases.

## Mean Square Displacement

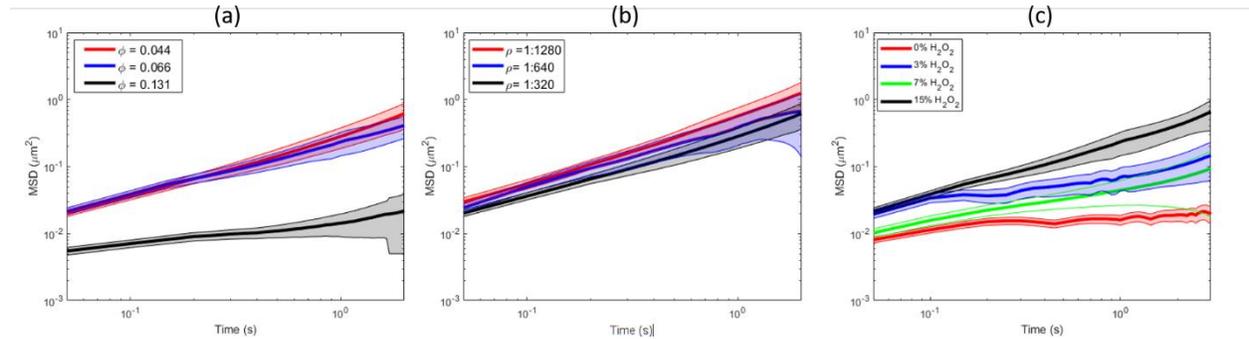

Figure S5: Mean square displacement of all particles in the colloidal crystal in the presence of active particles at different (a) surface densities (φ) at ρ=1:320 and 15% $H_2O_2$ (b) active particle ratios (ρ) at φ=0.044 and 15% $H_2O_2$ and (c) peroxide concentrations at (φ)=0.066 at ρ=1:320. The shaded boundaries indicate the standard deviation.

The mean square displacement (MSD) of the particles in the 2D colloidal crystal were measured for different compositions to explore the effect of defects on the mechanics of the 2D colloidal crystals. MSD, usually measured over a period of lag time, is a measure of particle displacement from its reference position. MSD can be related to the lag time (τ) as,

$$MSD = \beta \tau^\alpha \quad (1)$$

where, β is a factor related to the diffusion coefficient of the particles at the interface and exponent, α quantifies the mobility of the particles at the interface. For a viscoelastic interface, α varies between 0 and 1. α is zero for a purely elastic interface and 1 for a purely viscous interface. Figure S5(a) shows average MSDs of all the particles in the colloidal crystal at different φ values, in the presence of 3% $H_2O_2$ and ρ=1:320. With increasing surface density (φ), the slope of the MSDs, α reduces towards zero, indicating that the motion of the particles are restricted. However, at φ=0.044, the α value is close to 1, which indicates an increased viscous behavior of the colloidal crystal at lower surface density in the presence of activity. Likewise, an active particle ratio, ρ=1:1280 is sufficient to induce a viscous nature for the colloidal crystal at φ=0.044, as shown in Figure S5 (b). The effect of activity on the dynamics of the particles in the colloidal crystals can be clearly identified in Figure S5 (c) showing the effect of $H_2O_2$ concentration at φ=0.044 and ρ=1:320. In the presence of activity, the particles exhibit greater dynamics with slopes close to 1, while the slope remains close to 0 in the absence of activity (0% $H_2O_2$). In the case of lower surface density (φ=0.044), the peroxide concentration above 3% doesn't seem to affect the average MSD of the particles, as shown in Figure S5 (c). Hence, the creation of defects in the colloidal crystal due to the presence of activity increases the dynamics of the particles and causes a solid-to-liquid transition of the viscoelastic interface.

## Excess entropy calculations

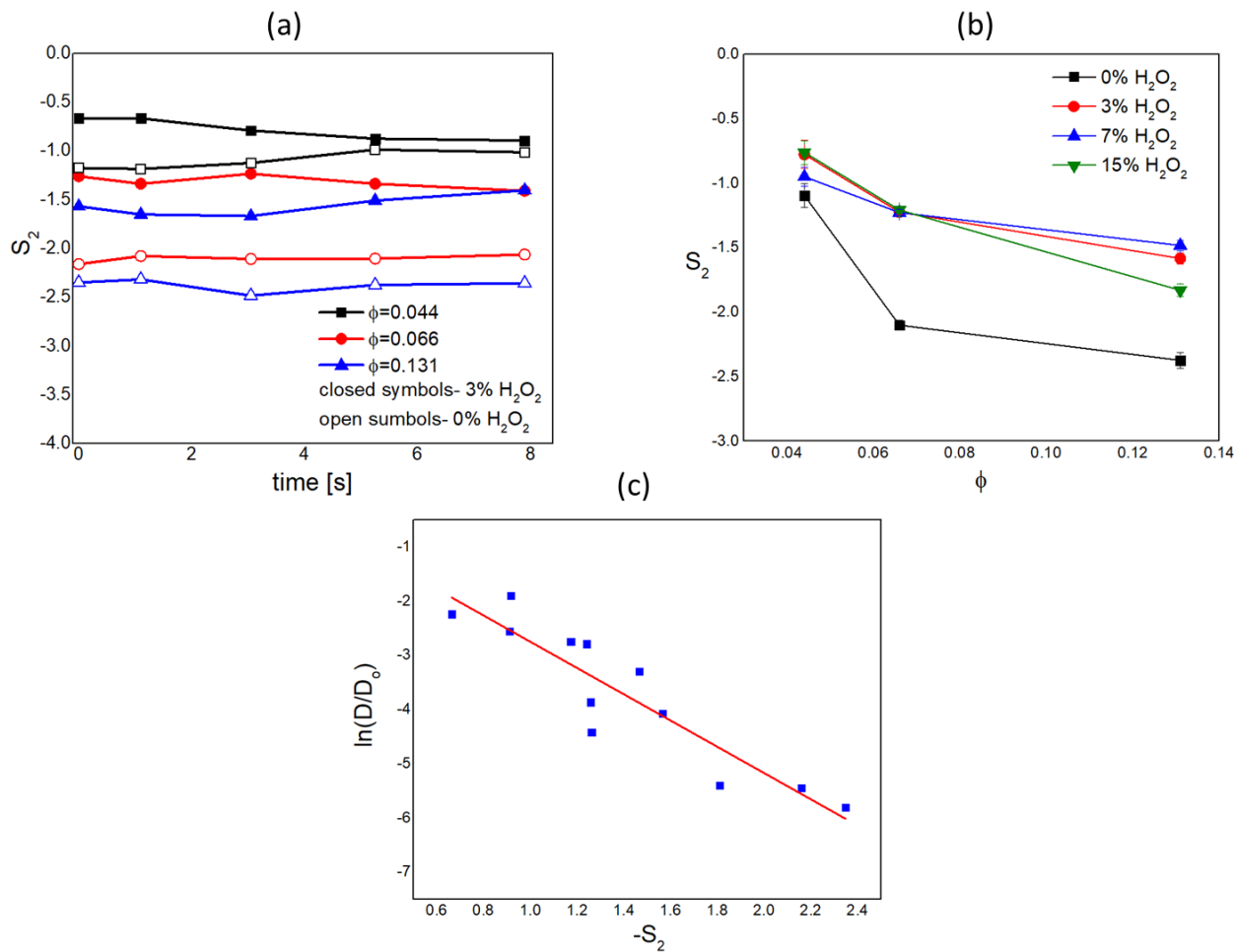

Figure S6: Excess entropy ($S_2$) measurements for 2D colloidal crystals with (a) ρ=1:320 and different φ values in the presence and absence of activity. (b) ρ=1:320, φ=0.044 and different $H_2O_2$ concentrations. (c) The variation of normalized diffusion coefficient ($D/D_o$) as a function of S2. The red solid line is the best exponential fit described as $D/D_o = e^{2.42 S2}$.

A characterization of the relationship between the microstructure and the macroscopic properties of the 2D colloidal crystals in the presence of activity can be performed by measuring the excess entropy ($S_2$) from the radial distribution function ($g(r)$) of the image microstructure[3]. $S_2$ is the difference between the system's entropy and the entropy of ideal gas ($S_2 = S - S_{IG}$). For 2D colloidal crystals, $S_2$ is a measure of the additional entropy associated with the organization of the colloidal particles in the crystal lattice. $S_2$ can be calculated from the $g(r)$ data using the expression,

$$S_2 = -\pi\rho \int_0^\infty g(r)\ln[g(r)] - [g(r) - 1] r \, dr \qquad \ldots\ldots\ldots(2)$$

where, ρ is the number density of the particles at the interface. We used Equation 2 to compute the excess entropy of the colloidal crystals in the presence of activity at different instances of time. In the case of 2D colloidal crystals, the particles are ordered and arranged in a specific lattice structure and hence its entropy is lower compared to a disordered state where the particles are randomly distributed. Figure S6 (a) shows the variation of $S_2$ as a function of time for the 2D colloidal crystals with different φ values, in the presence and absence of activity. The $S_2$ values are higher in the presence of activity for all the φ values, indicating an increase in disorder with activity. Additionally, the excess entropy decreases with increase in surface density, as shown in Figure S6 (b). Further, the normalized diffusion coefficient and $S_2$ exhibit an exponential scaling, $D/D_0 = e^{\alpha S_2}$ as proposed by Dzugutov.

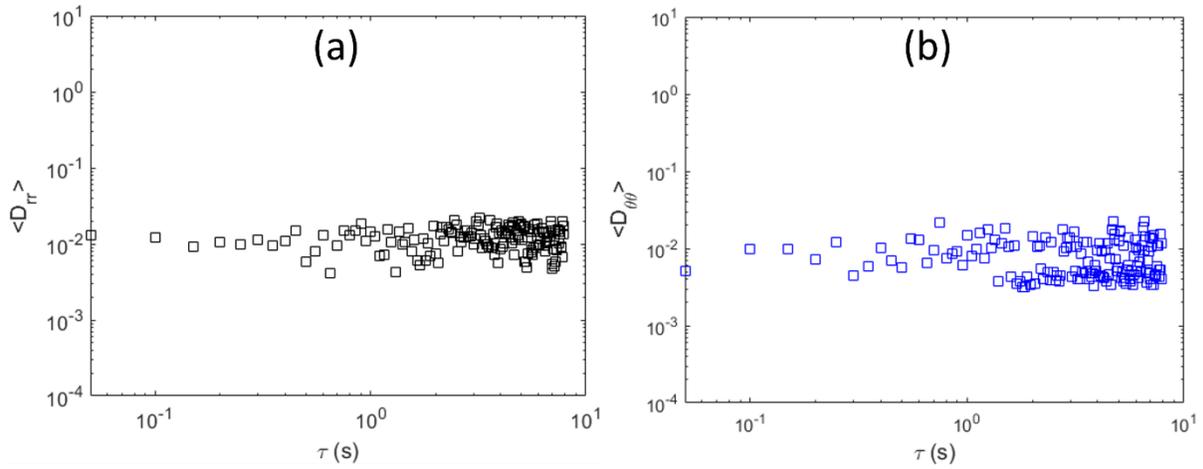

Figure S7: Variation of (a) $D_{rr}$ and (b) $D_{\theta\theta}$ as a function of τ at R=100 μm for φ=0.044, ρ=1:320 and 15% $H_2O_2$.

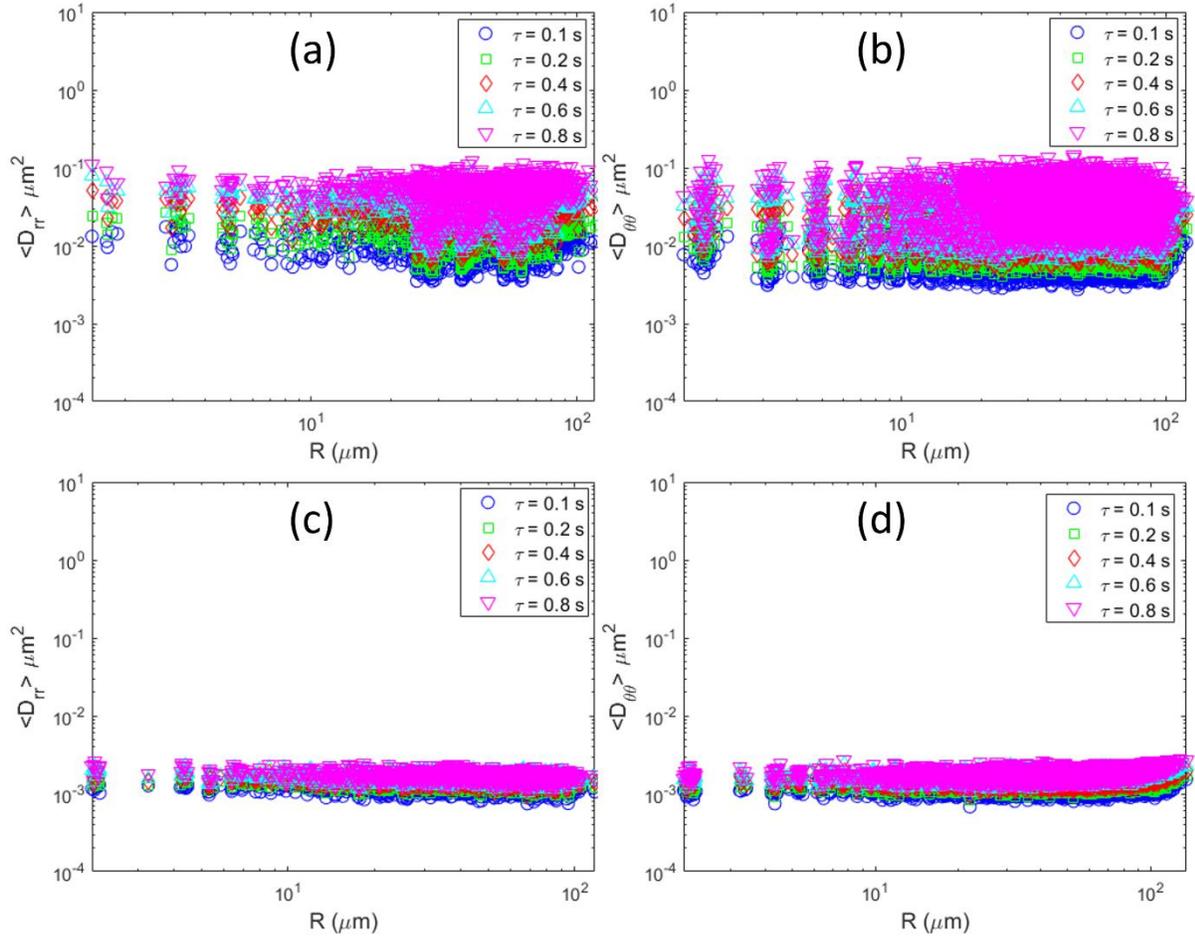

Figure S8: Variation of two-point correlation functions (a) $D_{rr}$ and (b) $D_{\theta\theta}$ for φ=0.044, 15% $H_2O_2$ and ρ=1:320. Similarly the variation (c) $D_{rr}$ and (b) $D_{\theta\theta}$ for φ=0.131, 0% $H_2O_2$ and ρ=1:320 at different lag times (τ).